\def\rn{\noindent\parshape 2 0truecm 8.8truecm 0.3truecm 8.5truecm}
\def\nn#1 #2{#1, #2.}				
\def\nnn#1 #2 #3{#1, #2. #3.}			
\def\nnnn#1 #2 #3 #4{#1, #2. #3. #4.}		
\def\nnnnn#1 #2 #3 #4 #5{#1, #2. #3. #4. #5.}	
\def\dualand{, \&\hbox{ }}				
\def\multiand{, \&\hbox{ }}				

\def\rg#1;#2;#3;#4;#5;#6 {\par\rn#1 #2, {\it #3}, {\bf #4}, #5 (``#6'') \par}
\def\rf#1;#2;#3;#4;#5 {\par\rn#1 #2, {\it #3}, {\bf #4}, #5\par}
\def\rfbook#1;#2;#3;#4;#5 {{\frenchspacing\par\rn#1 #2, {\it #3} (#4: #5)\par}}
\def\rfproc#1;#2;#3;#4;#5;#6 {{\frenchspacing\par\rn#1 #2, in {\it #3}, ed. #4 (#5: #6)\par}}
\def\rfprep#1;#2;#3  {{\par\rn#1 #2, #3\par}}
\def\rfprepp#1;#2;#3 {{\par\rn#1 #2, #3\par}}

\def\expec#1{\langle#1\rangle}

\def\etal{{\frenchspacing\it et al.}}

\def\eg{{\frenchspacing\it e.g.}}

\def\rms{rms}

\def\beq#1{\begin{equation}\label{#1}}
\def\eeq{\end{equation}}
\def\beqa#1{\begin{eqnarray}\label{#1}}
\def\eeqa{\end{eqnarray}}
\def\eq#1{equation~(\ref{#1})}

\def\eqn#1{~(\ref{#1})}

\def\spose#1{\hbox to 0pt{#1\hss}}
\def\simlt{\mathrel{\spose{\lower 3pt\hbox{$\mathchar"218$}}
     \raise 2.0pt\hbox{$\mathchar"13C$}}}
\def\simgt{\mathrel{\spose{\lower 3pt\hbox{$\mathchar"218$}}
     \raise 2.0pt\hbox{$\mathchar"13E$}}}
\def\simpropto{\mathrel{\spose{\lower 3pt\hbox{$\mathchar"218$}}
     \raise 2.0pt\hbox{$\propto$}}}

\def\ed{\end{document}}

\def\rt{\rho}
\def\rl{\rho_l}
\def\dt{\delta}
\def\dl{\delta_l}
\def\dperp{\delta_\bot}
\def\dtz{\delta_0}

\def\k{{\bf k}}
\def\r{{\bf r}}
\def\v{{\bf v}}
\def\x{{\bf x}}
\def\C{{\bf C}}
\def\xh{\widehat{\x}}
\def\Ch{\widehat{\C}}
\def\D{D}
\def\M{{\bf M}}
\def\Px{P_\times}

\documentstyle[emulateapj]{article}
\begin{document}


\journalid{337}{15 January 1989}
\articleid{11}{14}

\submitted{Submitted April 5 1998; accepted April 22; 
Published June 20 in {\it ApJL}, {\bf 500}, 79-82}

\title{THE TIME-EVOLUTION OF BIAS}

\author{
Max Tegmark
\footnote{Institute for Advanced Study, Princeton, 
NJ 08540; max@ias.edu}$^,$\footnote{Hubble Fellow}
and
P. J. E. Peebles\footnote{
Joseph Henry Laboratories, Princeton University, Princeton, NJ
08544; pjep@pupgg.princeton.edu} }

\begin{abstract}

We study the evolution of the bias factor $b$ and the mass-galaxy 
correlation coefficient $r$ in a simple
analytic model for galaxy formation and the gravitational growth
of clustering. The model shows that $b$ and $r$ can be strongly
time-dependent, but tend to approach unity even if
galaxy formation never ends as the gravitational growth of
clustering debiases the older galaxies. The presence of random
fluctuations in the sites of galaxy formation relative to the
mass distribution can cause large and rapidly falling bias values
at high redshift. 
 
\end{abstract}



\section{INTRODUCTION}

The relative distribution of galaxies and mass is of increasing
concern in cosmology. 
Constraints on cosmological parameters from galaxy surveys are only as accurate
as our understanding of bias.
Furthermore, if the density parameter
in mass that is capable of clustering were shown to be 
$\Omega =0.25\pm 0.15$ then most dynamical analyses of the
motions of galaxies would be
consistent with the assumption that galaxies trace mass, and the
challenge would be to explain why simulations of the
adiabatic cold dark matter (CDM) model for structure formation
typically indicate galaxies are more strongly clustered than
mass (Jenkins {\etal} 1998). If $\Omega=1$, galaxies do not trace mass and
it would be puzzling that the mass autocorrelation function in
CDM simulations is a much poorer approximation to a power law on
scales $10\hbox{ kpc}\simlt hr\simlt 10\hbox{ Mpc}$ than is the
galaxy two-point correlation function, and that the mass function
shows a much more pronounced evolution of shape back to 
$z\sim 1$ (Jenkins {\etal} 1998). 
Major surveys in progress of galaxies
and the cosmic microwave background (the CMB) will
advance our understanding of such issues. It may be
useful to supplement these observations with analytic
illustrations of how biasing can evolve as mass clustering grows
and gravity draws together the galaxies with the mass. The
purpose of this {\it Letter} is to present a simple analytic
model for the dynamical evolution of the relative distribution
of galaxies and mass. 

We assume that at formation a galaxy may be assigned a
near permanent observational tag. The far infrared luminosity
would not do, for IRAS galaxies avoid dense regions, an example
of biasing due to environment (\eg, Strauss \& Willick 1995).
The spheroid luminosity may be an adequate tag, for
the spheroid star populations in more luminous galaxies are
thought to be old and slowly evolving. We let $\rl(\r)$ be
the density of this luminous matter and $\rt(\r)$ be the total
mass density, both smoothed on some appropriate scale $R$. 
The corresponding density fluctuations are
$\dt\equiv\rt/\expec{\rt}-1$ and 
$\dl\equiv\rl/\expec{\rl}-1$. 
In a commonly used model $\delta_l = b\delta$, where the constant $b$ is
the bias factor. We adopt the more general and possibly more
realistic statistical representation of 
Dekel \& Lahav (1998; Dekel 1997 \S 5.5; Lahav 1996 \S 3.1) 
where $\delta$ and $\delta_l$ are
treated as 
stationary (translationally invariant) random processes
that may be grouped in the two-dimensional vector
\beq{xDefEq}
\x\equiv\left({\dt\atop\dl}\right).
\eeq
The mean $\expec{\x}$ vanishes by definition.
We write the covariance matrix as
\beq{Ceq}
\C\equiv\expec{\x\x^t}=\sigma^2
\left(\begin{tabular}{cc}
$1$&$br$\\
$br$&$b^2$
\end{tabular}\right),
\eeq
where $\sigma\equiv C_{11}^{1/2}$ is the 
{\rms} mass fluctuation on the scale $R$, 
$b\equiv(C_{22}/C_{11})^{1/2}$ is the bias factor
(the ratio of luminous and total fluctuations),
and $r\equiv C_{12}/(C_{11} C_{22})^{1/2}$ is the dimensionless 
correlation coefficient between the distributions of mass and
galaxies. If the smoothing scale $R$ is large enough that 
$\x$ can be modeled as a bivariate Gaussian random variable, 
then $\C$ contains all the statistical information about $\x$.
In principle this representation is not affected by galaxy
merging, if the process does not add much to the star population
that tags galaxies. In practice many of the surveys in progress
will use galaxy counts, but it is thought that there has not been
substantial merging of $L\simgt L_\ast$ galaxies since redshift
$z=1$. Pen (1998) has shown that the three quantities
$\sigma$, $b$ and $r$ can be measured 
from redshift space distortions (Kaiser 1987; reviewed
by Hamilton 1997). High redshift surveys in progress 
(see {\eg} Giavalisco {\etal} 1998; Yee {\etal} 1998
and the review in Moscardini {\etal} 1998) 
open the possibility of
measuring the time evolution of $b$ and $r$. Thus it seems 
timely to explore physical models for the evolution of these
second moment measures of biasing.

For studying the scale-dependence of bias, it is useful to 
work with $\xh$, the Fourier transform of $\x$,
which satisfies
$\expec{\xh(\k)\xh(\k')^\dagger}=(2\pi)^3\delta^D(\k-\k')\Ch(\k)$,
\beq{PdefEq2}
\Ch(\k )\equiv
\left(\begin{tabular}{cc}
$P(\k )$ & $\Px(\k )$ \\
$\Px (\k )$ & $P_l(\k )$
\end{tabular}\right) = P(\k)\left(\begin{tabular}{cc}
$1$ & $b'r'$ \\ $b'r'$ & $b'^2$ \end{tabular}\right).
\eeq
Here $P$ is the power spectrum of the mass distribution, 
$P_l$ is the power spectrum of the light that traces the
galaxies, and $\Px$ is the cross spectrum. The last expression
defines the analogs of the bias functions in \eq{Ceq}.
We have entered vector arguments to take
account of application in redshift space in fields of small
angular width. The following analysis applies equally well to $b$
and $r$ as to $b'$ and $r'$. 

Most previous analyses of biasing models have focused on
the static and local or non-local relation between 
$\delta_l$ and $\delta$ as galaxies form. Recent reviews 
are given by Mann {\etal} (1998), Croft {etal} (1998) 
and Scherrer \& Weinberg (1998);
the latter appears to be the first to make static model
predictions for $r$ as well as $b$. But these measures are
functions of time. Thus even if galaxies initially
were uncorrelated with the mass ($r=0$), they would gradually 
become correlated as gravity draws them towards
overdense regions, and one might expect 
this process to drive $b$ and $r$ toward unity.
Fry (1996) has demonstrated this explicitly 
for the special case $r=1$, and similar conclusions have
been found in numerical simulations of dark matter halo
clustering (\eg, Mo \& White 1996; Matarrese {\etal} 1997; 
Bagla 1998; Catelan 1998ab; Porciani 1998; Wechsler {\etal} 1998).

We limit our discussion to the linear perturbation theory regime  
$|\delta |\ll 1$, that is, to large scales $R$.
In \S 2 we derive a general expression 
for the time-evolution of $\C$ that applies
once galaxy formation has stopped. We generalize
to ongoing galaxy formation in
\S 3, and present conclusions in \S 4.

\section{AFTER THE GALAXY FORMATION EPOCH}

We assume galaxy formation
converts part of the mass into a near permanent luminous form 
without affecting its position or velocity on large scales, 
and that this luminous form behaves as test particles that are fair  
tracers of the large-scale velocity field (but not necessarily
the mass density). Since galaxies and mass are assumed to have 
the same bulk peculiar velocity $\v(\r)$, the contrasts $\dl$ and
$\dt$ satisfy the same linear continuity 
equation,  
$\dot\dl\approx\dot\dt\approx-\nabla\cdot\v$,
in the absence of galaxy formation. This gives
\beq{xEvolEq}
\x_0 = \M\x,\quad\hbox{where}\quad
\M \equiv 
\left(\begin{tabular}{cc}
$\D^{-1}$&$0$\\
$\D^{-1}-1$&$1$
\end{tabular}\right).
\eeq
The subscript indicates the present value, 
and $D$ is the growth factor of the mass density contrast in
linear perturbation theory (see {\eg} Peebles 1980)
normalized to the present value $\D(a_0)=1$.
The covariance matrix therefore evolves to
\beq{CevolEq}
\C_0 = \M\C\M^t.
\eeq
Equations \eqn{Ceq} to\eqn{CevolEq} give 
\beqa{CoolEq}
\sigma_0&=&\sigma/\D,\\
b_0     &=&[(1-\D)^2+2\D(1-\D) b r + \D^2 b^2]^{1/2},\label{bEq}\\
r_0     &=&[(1-\D) + \D b r]/b_0\label{req},
\eeqa
The first equation simply states that 
the total mass fluctuations have grown as 
$D(t)$. The second two equations show that {\it the situation tends
to grow simpler}, $b$ and $r$ approaching unity regardless of their
initial values. This is illustrated in Figure 1.
Inverting equations\eqn{CoolEq} to\eqn{req} gives
\goodbreak
\beqa{CoolEq2}
\sigma&=&\sigma_0\D,\\
b     &=&[(1-\D)^2 - 2(1-\D)b_0 r_0 + b_0^2]^{1/2}/\D,\label{bEq2}\\
r     &=&[b_0 r_0 - (1-\D)]/\D b\label{req2}.
\eeqa
This tells us what $\sigma$, $b$ and $r$ must have been in the past
to produce the present values.
In an Einstein-de Sitter universe ($\Omega_m=1$ and 
$\Omega_\Lambda=0$, as in the standard SCDM cold dark matter
model), the linear growth factor is  
$\D=a/a_0=(1+z)^{-1}$ and these equations are 
\beqa{CoolEq3}
\sigma&=&(1+z)^{-1}\sigma_0,\\
b     &=&[z^2 - 2z(1+z)b_0 r_0 + (1+z)^2 b_0^2]^{1/2},\label{bEq3}\\
r     &=&[(1+z)b_0 r_0 - z]/b\label{req3}.
\eeqa
For reference, these SCDM redshifts $z$
are given at the top of Figure 1. 

The value $r=1$ is a fixed point: if 
mass and galaxies are perfectly correlated at one
time this remains true for all time. 
For $r=1$, \eq{bEq} is
\beq{LinearSuppresionEq}
(b_0-1)=(b-1)\D,
\eeq
which for SCDM gives $b=b_0 + (b_0-1) z$, 
the special case derived by Fry (1996)
and Mo {\etal} (1997).
This is the top curve in each
set in Figure~1A. One also sees that if $r\sim 0$, 
$b$ tends to decrease even if there is no
initial bias. The correlation $r$ grows monotonically  
with time in all cases. The larger $b$, 
the slower the approach of $r$ to unity.

\centerline{{\vbox{\epsfxsize=9.0cm\epsfbox{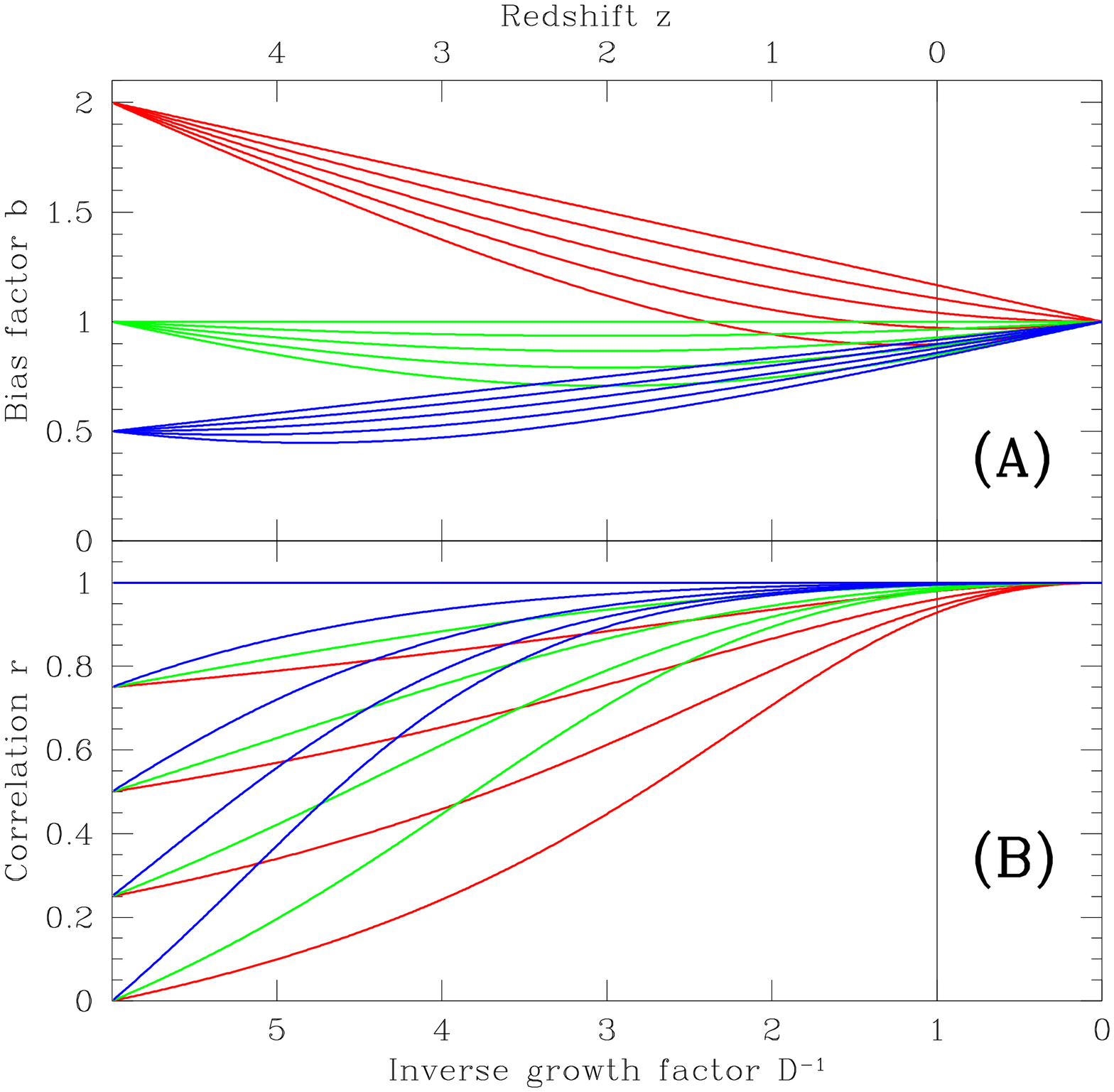}}}}
{\footnotesize {\bf FIG. 1}
--- The evolution of bias (A) and correlation
(B) is shown for 15 models that at 
Einstein-de Sitter redshift $z=5$ have 
bias $b=0.5$, 1, and 2 and correlation coefficients $r=0$, 0.25,
0.5, 0.75, and 1. In the top panel, $r$ increases upward 
in each quintuplet of lines. In the bottom panel, 
$b_0$ increases downward in each triplet of lines. The present
epoch is at $D^{-1}=1$, and the evolution is extrapolated to a
large future increase in the density contrast, $D^{-1}\rightarrow 0$.
The redshift scale at the top of the figure assumes the
Einstein-de~Sitter model.
}

\bigskip
\bigskip
\bigskip


\section{ONGOING GALAXY FORMATION}

We consider now the more general case where 
galaxies form over a substantial range of redshift. 
Here there is a competition between gravitational instability,
which pushes $b$ and $r$ toward unity, and the formation of
new galaxies, which can be biased or poorly correlated with 
the mass distribution. We model the time evolution of the
density of the galaxies, as measured by the luminosity tracer
$\rl$, as
\beq{GalFormEq}
\dot\rl+\rl\nabla\cdot\v = 
g(t)\left[1 + b_*\dt +\dperp\right],
\eeq
in comoving coordinates. 
The galaxy formation rate per unit volume is $g(t)$. 
The dimensionless factor in brackets models the sum of galaxy
formation determined by the local mass density and 
a random component $\dperp(\r,t)$. The latter has zero mean
($\expec{\dperp}=0$) and is uncorrelated with the mass density
($\expec{\dt\dperp}=0$). The deterministic part is represented as
the first two terms in a Taylor series expansion of the galaxy
formation rate as a function of the local mass density,
as in Coles (1993) and Fry \& Gazta\~naga (1993). The 
time-dependent parameter $b_*$ in this expansion is the 
``bias at birth''. 

As in the previous section, the streaming velocity of galaxies
and mass is related to the time-evolution of the mass contrast, 
$-\nabla\cdot\v=\dot\dt=\dot\D\dtz$.
The space average of \eq{GalFormEq} is to leading order
$\expec{\dot\rl}=g(t)$, so the mean galaxy density is
\beq{GdefEq}
\expec{\rl(t)}=G(t)\equiv\int_0^t g(t')dt'.
\eeq
Thus we can rewrite \eq{GalFormEq} as
\beq{GalFormEq2}
\dot\rl = g  + (b_*\D g + \dot\D G)\dtz + g\dperp ,
\eeq
and we integrate this to obtain $\rl$.
For any time-dependent quantity $f(t)$ 
we define the time average weighted by the galaxy formation rate
$g$ by 
\beq{AverageDefEq}
\expec{f}_t\equiv {1\over G(t)}\int_0^t f(t')g(t') dt'
	= {1\over G(t)}\int_0^{G(t)} fdG'.
\eeq
With this notation and the relation 
$\dot\D G = (\D G)\dot{} - \D g$, \eq{GalFormEq2} gives
the galaxy contrast 
\beq{dlEq}
\dl\equiv{\rl\over\expec{\rl}}-1=
c\dt  + \expec{\dperp}_t,
\eeq
where 
\beq{cDefEq}
c\equiv 1 + {\expec{(b_*-1)\D}_t\over\D}.
\eeq
Then the covariance matrix is 
\beq{Ceq2}
\C=\expec{\x\x^t}=\sigma^2
\left(\begin{tabular}{cc}
$1$&$c$\\
$c$&$c^2+s^2/\sigma^2$
\end{tabular}\right),
\eeq
where
\beq{sDefEq}
s^2\equiv
{1\over G^2}\int_0^G\int_0^G\expec{\dperp(G')\dperp(G'')}dG'dG'',
\eeq
in the notation of the last term in \eq{AverageDefEq}.
Combining \eq{Ceq} with \eq{Ceq2} yields
\beq{bEq4}
       b =[c^2+s^2/\sigma^2]^{1/2},\qquad r = c/b.\label{req4}
\eeq

Let us first consider some simple cases.
If there is no random contribution to galaxy formation,
$\dperp=0$, then 
$s^2=0$, $r=1$, and $b=c$.
If $r=1$ and the bias at birth, $b_*$, is time-independent, 
then \eq{cDefEq} reduces to $(b_0-1)=(b_*-1)\expec{\D}_t$. 
Here the deviation of the bias from unity is suppressed 
by the average growth factor between the galaxy formation epoch 
and today.
This is just \eq{LinearSuppresionEq}
averaged over the galaxy formation history.
In the the Einstein-de~Sitter model this bias suppression factor 
$\expec{\D}_t$ is $\expec{(1+z)^{-1}}_t$.

\bigskip

{\footnotesize
{\bf Table 1} -- Resulting $b_0$ and $r_0$ for various models,
for a constant galaxy formation rate $g(z)$ between 
redshifts $z_{on}$ and $z_{off}$.

\noindent
\begin{tabular}{|c|cc|cc|ccc|cc|}
\hline
Model	&$\Omega_0$	&$\Lambda_0$	&$z_{on}$	&$z_{off}$&$b_*$	&$s_*$	&$N$	&$b_0$	&$r_0$\\
\hline
M1	&1	&0	&5	&5	&2	&0	&1	&1.17	&1\\	
M2	&0.3	&0.7	&5	&5	&2	&0	&1	&1.21	&1\\	
M3	&0.3	&0	&5	&5	&2	&0	&1	&1.30	&1\\	
\hline
M4	&1	&0	&5	&0	&2	&0	&1	&1.36	&1\\	
M5	&1	&0	&1	&0	&2	&0	&1	&1.69	&1\\	
M6	&1	&0	&3	&1	&2	&0	&1	&1.35	&1\\	
M7	&1	&0	&5	&2	&2	&0	&1	&1.23	&1\\	
M8	&1	&0	&7	&3	&2	&0	&1	&1.17	&1\\	
\hline
M9	&1	&0	&5	&2	&0.5	&0	&1	&0.88	&1\\	
M10	&1	&0	&5	&2	&1	&0	&1	&1	&1\\	
\hline
M11&1	&0	&5	&2	&1	&0.3	&1	&1.08	&.93\\	
M12&1	&0	&5	&2	&1	&0.3	&10	&1.01	&.99\\	
M13&1	&0	&5	&2	&2	&0.3	&1	&1.29	&.95\\	
M14&1	&0	&5	&2	&2	&0.3	&10	&1.24	&.99\\	
\hline
\end{tabular}
}

\bigskip

\centerline{{\vbox{\epsfxsize=9.0cm\epsfbox{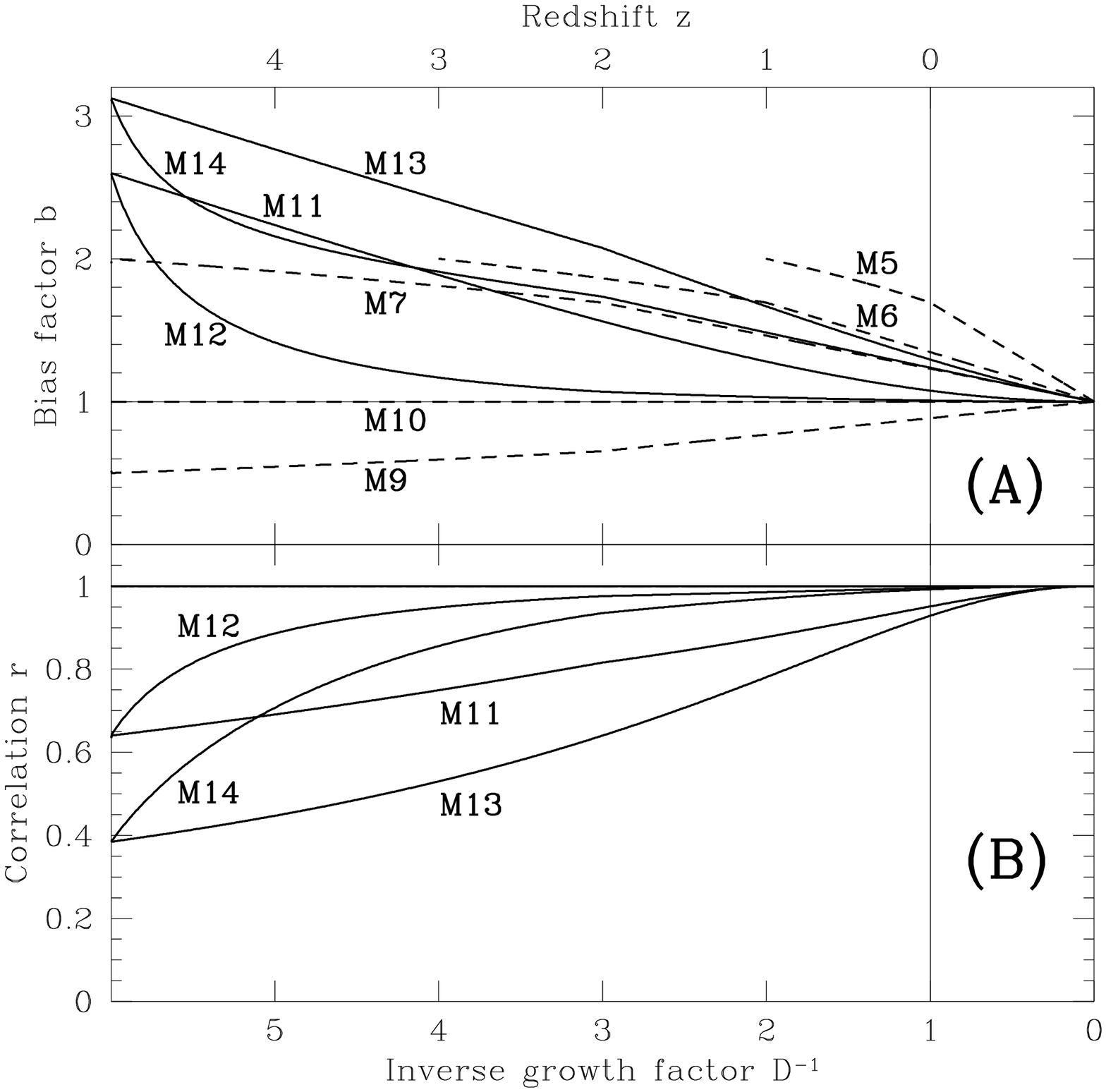}}}}
{\footnotesize {\bf FIG. 2}
--- The time-evolution of bias (A) and correlation
(B) for some of the models in Table 1.
}

\bigskip

The randomness contribution $s^2$ to the correlation $r$ 
depends on $\expec{\dperp(t')\dperp(t'')}$, 
which is a measure of the memory of fluctuations 
$\dperp$ at the same comoving position at 
different times. If memory were complete, 
$\dperp$ a function of position alone, 
then $s=\expec{\dperp^2}^{1/2}_t$. In the limit of a short coherence time, 
$\expec{\dperp(G')\dperp(G'')}\propto\delta(G'-G'')$, 
\eq{sDefEq} gives $s\propto G^{-1/2}$, just as 
the average of $N$ independent fluctuations is proportional to 
$N^{-1/2}$. The simple two-parameter model
\beq{sModelEq}
s = {s_*\sigma_0\over\sqrt{1+(N-1){G\over G(a_0)}}} 
\eeq
incorporates both of these extremes as special cases
($N=1$ and $N=\infty$, respectively), and 
since this aspect of the galaxy formation process is still so poorly
understood a more complicated model for $s$
does not yet seem warranted. 
The parameter $N$ can be interpreted 
as the effective number of uncorrelated galaxy formation epochs.
The current rms mass contrast is $\sigma_0$, 
and the other model parameter is the relative normalization 
$s_*$.

Examples of these relations are listed in Table 1 and plotted in
Figure 2. The growth factors $\D$ for the low density models are
computed as in Carroll {\etal} (1992). These models show that the
biasing parameters $r$ and $b$ can vary quite rapidly near the
start of galaxy formation, and that $r$ and $b$ tend to unity
even if galaxy formation never ends, as in models 4 and 5,
because the growth of gravitational clustering debiases the
growing number of older galaxies. After galaxy formation
terminates  $b$ and $r$ approach unity 
more rapidly, evolving as in Figure 1. A random component
$\dperp$ produces a large early value of $b$,
but $b$ and $r$ approach unity quite rapidly if $N\gg 1$, as the
fluctuations from large numbers of random events average down to
insignificant levels.

\section{CONCLUSIONS}

We have computed the evolution of the bias factor $b$ and the
mass-galaxy correlation coefficient $r$ in a simple analytic
model represented by a galaxy formation history $g(z)$,
the parameters in a representation of
biased galaxy formation ($b_*$, $s_*$, and $N$), 
and the parameters for the cosmological model
($\Omega$ and $\Lambda$). Galaxy formation could be
affected by explosions, as from quasars, at positions 
unrelated to the local mass density (\eg, 
Dekel \& Rees 1987; Babul \& White 1991). 
This effect is represented by the term $\dperp$. The 
coherence time of $\dperp$, as measured by $N$, would depend on
how long individual quasars last and how long the intergalactic
medium ``remembers'' their effects. The 
complications of luminosity evolution
and galaxy mergers are incorporated into our formalism by
defining $g(z)$ to be the formation history of the matter
that is luminous at the redshift $z'$ at which we wish to evaluate
$b$ and $r$.

If galaxies were assembled at $z\simlt 1$ there could be a large
and observationally significant evolution of the biasing
functions $b$ and $r$ at low redshift, as illustrated by the
steep initial slopes in Figure~2. On the other hand, if galaxy
positions were assigned at relatively high redshift the effect of
biasing at birth would 
be strongly suppressed. Decreasing the mass density $\Omega$
decreases this effect, because the growth of density
perturbations at low redshifts is slower, but this likely is at
least partially canceled by shift of the formation epoch to
larger redshifts. In fact, if $g$, 
$b_*$, and $\dperp$ were
functions of $D$ alone,
debiasing would be independent of the cosmology. 

The new galaxy redshift surveys, including
the two degree field (2dF) survey and the 
Sloan Digital Sky Survey,
have the potential 
to measure key cosmological parameters with 
great accuracy, both alone
(Tegmark 1997; Goldberg \& Strauss 1998), 
and when combined with cosmic microwave background
experiments (Hu {\etal} 1997).
To achieve this, however, 
biasing and its evolution must be understood to a 
comparable accuracy. Our formalism deals with the second moments
of the mass and galaxy distributions in a universe that is
statistically homogeneous (eq.~[\ref{PdefEq2}]).
The trend to debiasing of the functions $b$ and $r$ in this
representation appears to be a generic feature of the
gravitational instability scenario for structure formation,
a prediction that may be observationally tested in the near
future as galaxy redshift surveys improve.

\smallskip
We thank Ben Bromley and Ue-Li Pen for useful 
discussions. Support for this work was provided by
NASA though grant NAG5-6034 and 
a Hubble Fellowship,
HF-01084.01-96A, awarded by STScI, which is operated by AURA, Inc. 
under NASA contract NAS5-26555, and at Princeton University by
the National Science Foundation. 



\bigskip
\bigskip

This paper is available with figures and links from 
{\it h t t p://www.sns.ias.edu/$\tilde{~}$max/bias.html}


\begin{references}   %

\rf\nn Babul A\dualand\nnnn White S D M;1991;MNRAS;253;31P

\rfprep\nnn Bagla J S;1998;astro-ph/9711081


\rfprep\nn Catelan P, \nn Lucchin  F, \nn Matarrese S\multiand\nn Porciani C;1998a;astro-ph/9708067
version matches the refereed one, in press in MNRAS       
       
\rfprep\nn Catelan P, \nn Matarrese S\multiand\nn Porciani C;1998b;astro-ph/9804250

\rf\nnn Carroll S M, \nnn Press W H\multiand\nnn Turner E L;1992;ARA\&A;30;499

\rf\nn Coles P;1993;MNRAS;262;1065

\rfprep\nn Croft R, \nn Dalton G\multiand\nn Efstathiou G;1998;astro-ph/9801254

\rfprep\nn Dekel A;1997;astro-ph/9705033



\rn\nn Dekel A\dualand\nn Lahav O 1998, in preparation

\rf\nn Dekel A\dualand\nnn Rees M J;1987;Nature;326;455

      
\rf\nn Fry J N;1996;ApJ;461;L65


\rf\nnn Fry J N\dualand Gazta\~naga E;1993;ApJ;413;447

      
\rfprep\nn Giavalisco M {\etal};1998;astro-ph/9802318

\rf\nnn Goldberg D M\dualand\nnn Strauss M A;1998;ApJ;495;;29

\rfprepp\nnnn Hamilton A J S;1997;astro-ph/9708102

\rfprep\nn Hu W, \nnn Eisenstein D J\multiand\nn Tegmark M;1998;astro-ph/9712057

\rf\nn Jenkins A {\etal};1998;ApJ;499;20
\rf\nn Kaiser N;1987;MNRAS;227;1


\rf\nn Lahav O;1996;Helvetica Physica Acta;69;{388, astro-ph/9611093}

\rf\nn Mann B, \nn Peacock J\multiand Heavens A;1998;MNRAS;293;209

\rf\nn Matarrese S, \nn Coles P, 
\nn Lucchin F\multiand\nn Moscardini L;1997;MNRAS;286;115

\rf\nnn Mo H J\dualand\nnnn White S D M;1996;MNRAS;282;347

\rf\nnn Mo H J, \nnn Jing Y P\multiand\nnnn White S D M;1997;MNRAS;284;189 
      
\rfprep\nn Moscardini L, \nn Coles P, 
\nn Lucchin F\multiand\nn Matarrese S;1997;astro-ph/9712184

\rn Peebles, P. J. E 1980, {\it The Large-Scale Structure of the Universe}
(Princeton U. P., Princeton)

\rfprep\nn Pen U;1998;astro-ph/971117

\rfprep\nn Porciani C, \nn Matarrese S, Lucchin F\multiand\nn Catelan P;1998;astro-ph/9801290
       
\rfprep\nnn Scherrer R J\dualand\nnn Weinberg D H;1998;astro-ph/9712192

\rf\nnn Strauss M A\dualand\nn Willick G;1995;Phys. Rep.;261;271

\rf\nn Tegmark M;1997;Phys. Rev. Lett.;79;3806

\rfprep\nnn Wechsler R H {\etal};1998;astro-ph/9712141

\rfprep\nnnnn Yee H K C {\etal};1998;astro-ph/9710356

\end{references}
\end{document}